\documentclass[letter]{jpsj3}
\title{Superconductivity Induced by Bond Breaking \\ in the Triangular Lattice of IrTe$_2$}

\author{
Sunseng \textsc{Pyon}$^{1,2}$\thanks{E-mail address : pyon@science.okayama-u.ac.jp}, Kazutaka \textsc{Kudo}$^{1,2}$, and Minoru \textsc{Nohara}$^{1,2}$
}
\inst{\address{2-31-22-5F Yushima, Bunkyo-ku, Tokyo 113-0034}
}

\inst{$^1$Department of Physics, Faculty of Science, Okayama University, Okayama  700-8530, Japan \\
$^2$Transformative Research-Project on Iron Pnictides (TRIP), Japan Science and Technology Agency (JST), 5 Sanbancho, Chiyoda-ku, Tokyo 102-0075, Japan}

\abst{
IrTe$_2$, a layered compound with a triangular iridium lattice, exhibits a structural phase transition at approximately 250 K. 
This transition is characterized by the formation of Ir-Ir bonds along the $b$-axis. 
We found that the breaking of Ir-Ir bonds that occurs in Ir$_{1-x}$Pt$_x$Te$_2$ results in the appearance of a structural critical point in the $T$ = 0 limit at $x_c$ $\simeq$ 0.035. 
Although both IrTe$_2$ and PtTe$_2$ are paramagnetic metals, superconductivity at $T_c$ = 3.1 K is induced by the bond breaking in a narrow range of $x \geq x_c$ in Ir$_{1-x}$Pt$_x$Te$_2$. 
This result indicates that structural fluctuations can be involved in the emergence of superconductivity.
}

\kword{superconductivity, IrTe$_2$, Pt doping, triangular lattice, phase diagram, structural phase transition}

\begin{document}

\maketitle

Strongly correlated electron systems exhibit a rich variety of electronic phases. 
When two phases meet each other in the $T$ = 0 limit, quantum fluctuations often give rise to exotic electronic states. 
The most intensively studied phenomenon is the breakdown of the Fermi liquid at magnetic quantum critical points (QCPs) in itinerant magnets, which include heavy fermion systems, transition-metal oxides and chalcogenides, and organic molecular conductors \cite{RevModPhys.73.797}. 
One question that arises is whether similar QCPs and exotic superconductivity can occur in systems other than itinerant magnets. 
Intriguing candidates for exhibiting such behavior are transition-metal compounds with geometrically frustrated lattices, such as triangular or pyrochlore lattices, which often form a valence-bond solid (VBS) state at low temperatures. 
When the $t_{2g}$ orbitals are partially occupied, using the orbital degrees of freedom, complex ``molecular" clusters with spin-singlet bonds are often formed: the linear chain in NaTiO$_2$ \cite{JPSJ.61.2156, doi:10.1021/cm970538c}, the trimer in LiVO$_2$ \cite{Imai1995184} and LiVS$_2$ \cite{PhysRevLett.103.146405}, the heptamer in AlV$_2$O$_4$ \cite{PhysRevLett.96.086406}, the helical dimer in MgTi$_2$O$_4$ \cite{PhysRevLett.92.056402}, and the octamer in CuIr$_2$S$_4$ \cite{PB194-196(1994)1077,Nature.416.155}. 
A VBS state has also been found in organic systems. 
Shimizu {\it et al.} demonstrated that the breaking of the VBS is possible by applying hydrostatic pressure to the organic triangular lattice system EtMe$_3$P[Pd(dmit)$_2$]$_2$ \cite{PhysRevLett.99.256403}. 
Remarkably, superconductivity appears at the critical point of the VBS transition. 
No inorganic systems have been reported to exhibit superconductivity at the VBS critical point, where bond breaking takes place.

IrTe$_2$, which has a triangular iridium lattice, exhibits a structural phase transition analogous to that of NaTiO$_2$. 
In this Letter, we demonstrate that the breaking of Ir-Ir bonds occurs for the solid solution Ir$_{1-x}$Pt$_x$Te$_2$ at $x_c$ = 0.035; this results in the appearance of a structural critical point at the onset of the structural phase transition in the $T$ = 0 limit. 
Although both IrTe$_2$ and PtTe$_2$ are paramagnetic metals, superconductivity at a maximum $T_c$ of 3.1 K emerges in the vicinity of the structural critical point at $x_c$\cite{NS2_pyon,ICNSCT_pyon}. 
We propose that structural fluctuations, likely to be related to the iridium $t_{2g}$ orbitals, can be involved in the occurrence of superconductivity.

IrTe$_2$ and PtTe$_2$ crystallize in a trigonal CdI$_2$-type structure with the space group $P\bar{3}m1$ ($\sharp$ 164). 
The edge-sharing IrTe$_6$ (PtTe$_6$) octahedra are in IrTe$_2$ (PtTe$_2$) layers, which are bound to each other by van der Waals forces. 
Each of these layers forms a regular triangular lattice of Ir (Pt) ions with three equivalent Ir-Ir (Pt-Pt) bonds, as depicted in the inset of Fig.~1(b). 
PtTe$_2$ is a simple metal down to the lowest temperature measured, without any trace of anomalies \cite{IrTe2_Matsumoto}. 
In contrast, IrTe$_2$ exhibits a first-order structural phase transition from the trigonal structure to a monoclinic structure with space group $C2/m$ ($\sharp$ 12) at approximately 250 K \cite{IrTe2_Matsumoto}. 
The transition is accompanied by both a jump in electrical resistivity and a discontinuous decrease in Pauli paramagnetic susceptibility, as shown in Fig.~2. 
This type of anomaly is often observed in charge-density wave (CDW) materials. 
However, modulation of the crystal lattice, a hallmark of CDW materials, has not been observed in IrTe$_2$. 
Instead, the transition is accompanied by the formation of ``uniform" Ir-Ir bonds along one side of the triangular lattice, as depicted in the inset of Fig.~1(b), while the remaining two Ir-Ir bonds are essentially unchanged by the transition \cite{IrTe2_Matsumoto}. 
Thus, the transition can be characterized by the formation of Ir-Ir bonds along the $b$-axis of the monoclinic unit cell, which deforms the regular triangular lattice into an isosceles triangle. 
No superconductivity has been reported for IrTe$_2$ down to 0.32 K \cite{Raub19652051} or for PtTe$_2$ down to 1.2 K \cite{RevModPhys.35.1}.

Polycrystalline samples of solid solutions of Ir$_{1-x}$Pt$_x$Te$_2$ with 0.00 $\leq$ $x$ $\leq$ 0.25 were synthesized using a solid-state reaction. 
Stoichiometric amounts of Ir, Pt, and Te were mixed, pelletized, and sealed in an evacuated quartz ampule. 
The ampule was heated at 900 $^\circ$C for 24 h. 
After cooling to room temperature over 24 h, the product was ground, pelletized, and heated at 900  $^\circ$C for 24 h in an evacuated quartz ampule. 
The products were characterized by powder X-ray diffraction and confirmed to consist of a single phase with a negligible amount of nonreacted Pt and Ir. 
Magnetization $M$ was measured using a SQUID magnetometer (Magnetic Property Measurement System, Quantum Design). 
Electrical resistivity $\rho$ was measured by the standard DC four-terminal method using a Physical Property Measurement System (PPMS, Quantum Design). 
Specific heat $C_p$ was measured by the relaxation method using the PPMS.

Figure~1(a) shows the lattice parameters $a$ and $c$ (in the hexagonal setting) and the unit cell volume $V_{\rm cell}$ for Ir$_{1-x}$Pt$_x$Te$_2$ as a function of $x$ at room temperature. 
The monotonic changes with Pt doping indicate the formation of a solid solution between IrTe$_2$ and PtTe$_2$. 
The trigonal $a$ parameter, which corresponds to the bond length between transition-metal ions (TM-TM bonds, TM = either Ir or Pt), increases from  3.932(1) {\AA} for $x$ = 0 to 3.964(1) {\AA} for $x$ = 0.25. 
This result suggests that Pt doping breaks the TM-TM bonds, which are formed at temperatures below approximately 250 K in IrTe$_2$.

\begin{figure}
\begin{center}
\includegraphics[width=8cm]{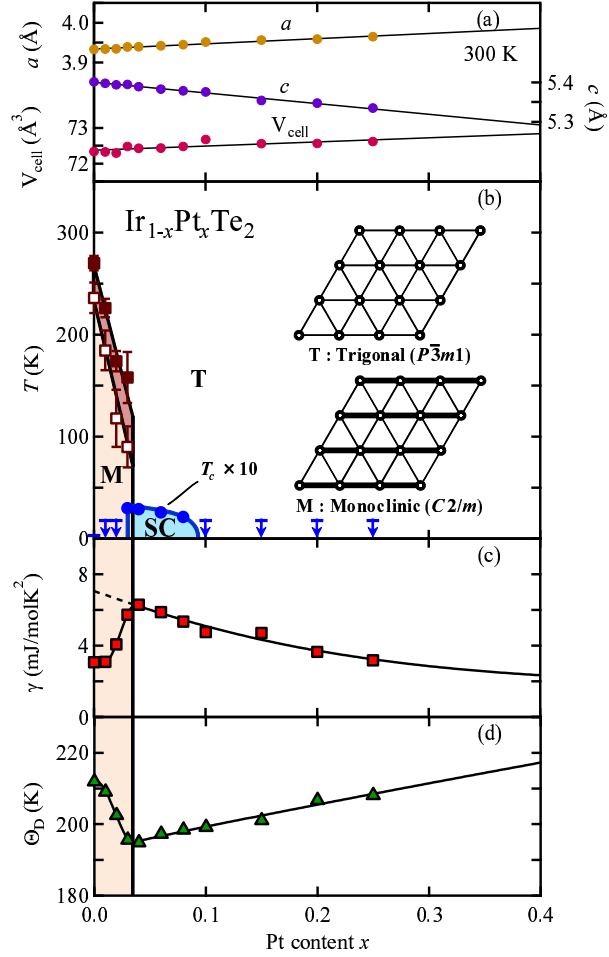}
\caption{\label{fig1}
(Color online)
(a) Lattice parameters and unit-cell volume in the trigonal lattice at room temperature as a function of $x$ for Ir$_{1-x}$Pt$_x$Te$_2$.  
(b) Electronic phase diagram of Ir$_{1-x}$Pt$_x$Te$_2$ versus Pt content $x$. T, M, and SC denote the trigonal phase, monoclinic phase, and superconducting phase, respectively. Closed circles show the superconducting transition temperature $T_c$ determined from specific-heat measurements. Bars and arrows indicate the absence of bulk superconductivity above 2.0 K at $x$ $>$ 0 and above 0.32 K at $x$ = 0. Closed and open squares show the structural transition temperature $T_s$ determined from transport measurements upon heating and cooling, respectively. The inset schematically illustrates the regular triangular lattice of the trigonal phase and the isosceles triangular lattice of the monoclinic phase. The thick, solid lines represent short chemical bonds, whereas the thin, solid lines represent long chemical bonds.
(c) Electronic specific-heat coefficient $\gamma$ and (d) Debye temperature $\Theta_D$ in the low-temperature limit as a function of $x$. 
}
\end{center}
\end{figure}

\begin{figure}
\begin{center}
\includegraphics[width=8cm]{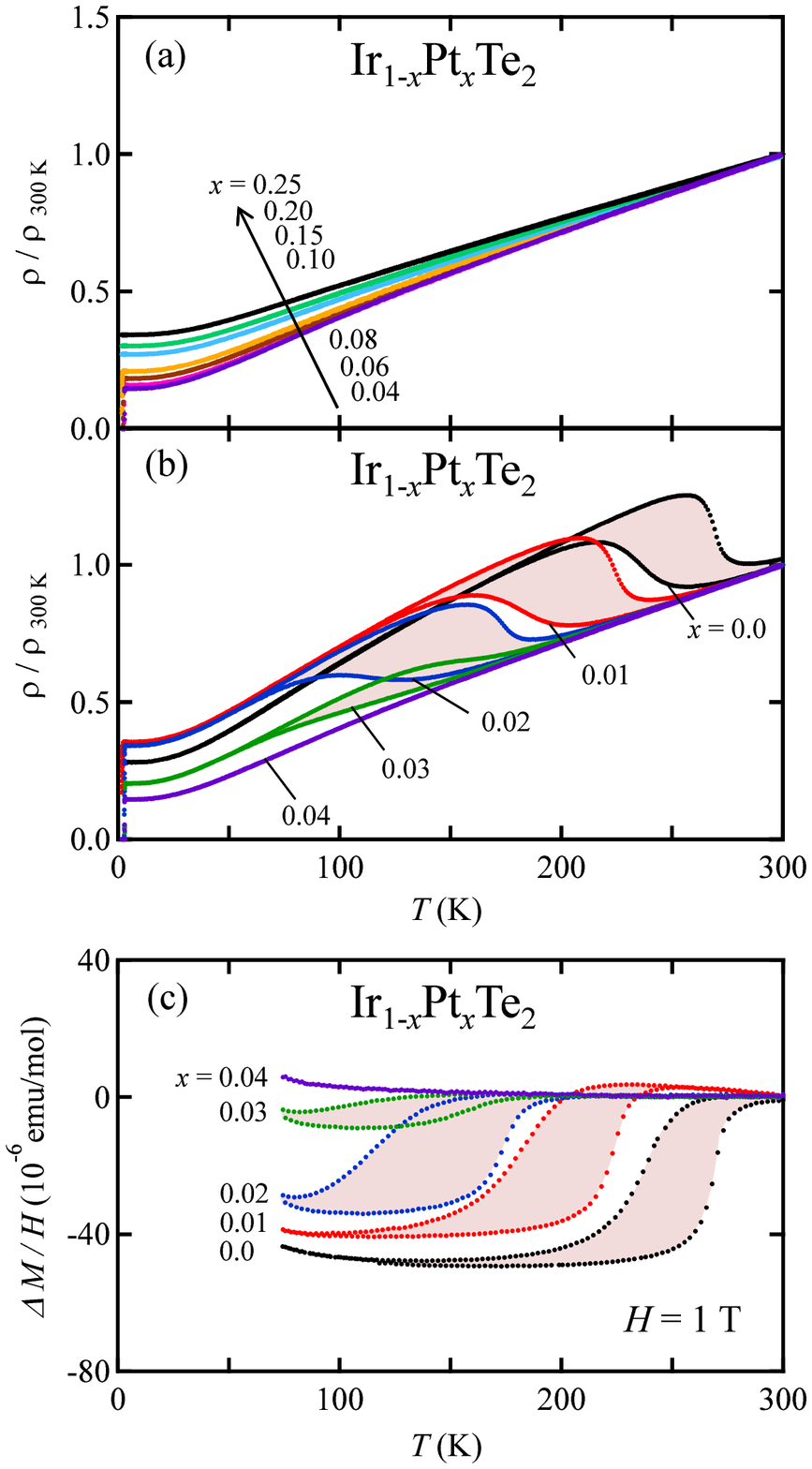}
\caption{\label{fig2}
(Color online)
(a), (b) Electrical resistivity $\rho$ of Ir$_{1-x}$Pt$_x$Te$_2$ (normalized by the value of $\rho$ at room temperature) as a function of temperature ((a) 0.04 $\leq$ $x$ $\leq$ 0.25, (b) 0.0 $\leq$ $x$ $\leq$ 0.04).
(c) Magnetic susceptibility measured at $H$ = 1 T as a function of temperature. 
The value at 300 K is subtracted for clarity ($\Delta M = M - M_{300 \mathrm{K}}$).
}
\end{center}
\end{figure}

The structural transition temperature $T_s$ can be determined from electrical and magnetic measurements. 
Figures 2(a) and 2(b) show the electrical resistivity $\rho$ divided by its value at room temperature for Ir$_{1-x}$Pt$_x$Te$_2$. 
The residual resistivity was in the range of 10 to 35 $\mu\Omega$cm, indicating that the system is a good metal. 
A systematic change occurs upon increasing $x$. $\rho$ shows clear anomalies at the structural phase transitions in the range of
 $x$ $\leq$ 0.03, very similar to those previously reported for pure IrTe$_2$ \cite{IrTe2_Matsumoto}. 
At the same temperature, magnetic susceptibility exhibits a jump, as shown in Figure 2(c). 
The transition is accompanied by thermal hysteresis, suggesting a first-order transition. 
The transition temperature $T_s$ decreases with increasing Pt doping and is suppressed completely at $x$ = 0.04, thereby indicating a critical concentration of $x_c$ $\simeq$ 0.035.

As the structural phase transition is suppressed with increasing Pt content, the superconducting phase emerges at low temperatures for 0.03 $\leq$ $x$ $<$ 0.10. 
This is illustrated in the low-temperature resistivity and magnetization data shown in Figs.~3(a) and 3(b), respectively. 
No superconducting transition was observed in pure IrTe$_2$. 
At $x$ = 0.01, however, a drop in the resistivity begins to appear. 
The superconducting transition is clearly observed for $x$ = 0.02 in both the resistivity and magnetization data. 
However, the small specific-heat jump at $T_c$, shown in Fig.~3(c), suggests that the superconductivity observed for $x$ = 0.02 is not a bulk property. 
$T_c$ reaches a maximum of about 3.1 K for $x$ = 0.03 and then decreases at higher Pt contents. 
Correspondingly, specific heat shows a clear peak at $T_c$, a hallmark of bulk superconductivity. 
No trace of superconductivity was observed for $x$ $\geq$ 0.10 down to 2.0 K. 
Thus, the optimal composition for the superconductivity of Ir$_{1-x}$Pt$_x$Te$_2$ with $x$ = 0.03 is observed at the critical boundary of the structural phase transition at the $T$ = 0 limit. 

\begin{figure}[h]
\begin{center}
\includegraphics[width=8cm]{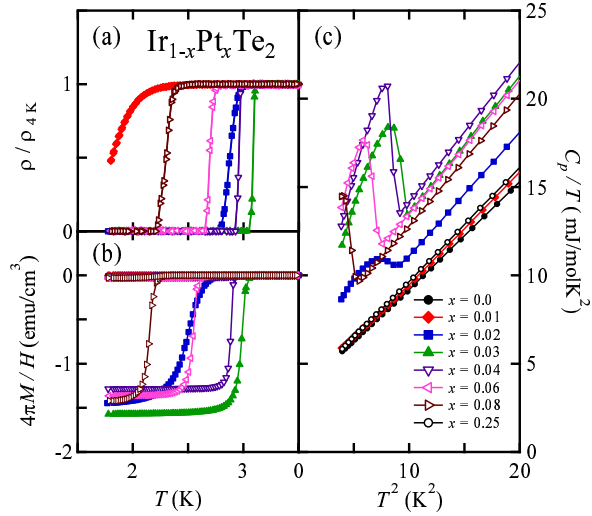}
\caption{\label{fig3}
(Color online)
(a) Low-temperature resistivity of Ir$_{1-x}$Pt$_x$Te$_2$ in zero applied field. 
(b) Low-temperature magnetization of Ir$_{1-x}$Pt$_x$Te$_2$. The measurements were conducted in an applied field of $H$ = 10 Oe with zero-field-cooled (ZFC) and field-cooled (FC) processes. The shielding volume fraction was nearly 100\% for 0.02 $\leq$ $x$ $\leq$ 0.08. 
(c) Specific heat divided by temperature $C_p/T$ as a function of $T^2$ in zero applied field. 
}
\end{center}
\end{figure}

Specific heat $C_p$ is shown in Fig.~3(c) in the form of a $C_p/T$ versus $T^2$ plot. 
The normal-state specific heat can be approximated at low temperatures by $C_p = \gamma T + \beta T^3$, where $\gamma$ represents the normal-state electron contribution and $\beta T^3$ represents the phonon contribution to the specific heat. 
The data show linear behavior in the $C_p/T$ versus $T^2$ plot between $T_c$ and 10 K; the extrapolation to $T$ = 0 and the slope give an estimate of $\gamma$ and the Debye temperature $\Theta_D$, as summarized in Figs.~1(c) and 1(d), respectively. 
In the trigonal phase for $x$ $\geq$ 0.04, $\gamma$ monotonically decreases with increasing $x$. 
This behavior is qualitatively consistent with a band calculation in which the electronic density of states (DOS) of IrTe$_2$ decreases above the Fermi level \cite{Soulard20052008}. 
The partial substitution of Pt for Ir has the effect of doping electrons and shifting the Fermi level upward, thus leading to a reduction of the DOS.
We estimate $\gamma$ to be approximately 7 mJ/molK$^2$ for pure IrTe$_2$ in the high-temperature trigonal phase by extrapolating the $\gamma$ versus $x$ curve to $x$ = 0, as shown by the broken line in Fig.~1(c). 
The estimated value of $\gamma$ is larger than that of $\gamma_{\rm band}$ = 5.1 mJ/molK$^2$ predicted using a band calculation \cite{Soulard20052008}, which is indicative of mass enhancement. 
This is in accordance with the Wilson ratio of $R_{\rm W}$ $\simeq$ 1.6 (larger than unity) calculated from $\gamma$ $\simeq$ 7 mJ/molK$^2$ and the Pauli paramagnetic susceptibility of $\chi_{\rm Pauli}$ $\simeq$ 1.5 $\times$ 10$^{-4}$ emu/mol after subtracting the core diamagnetism \cite{IrTe2_Matsumoto}.

The specific-heat coefficient $\gamma$, as well as the Pauli susceptibility, is reduced in the monoclinic phase for $x$ $\leq$ 0.03. 
The $\gamma$ and $\chi_{\rm Pauli}$ values for pure IrTe$_2$ in the monoclinic phase are approximately half of those expected in the trigonal phase, suggesting that a large part of the Fermi surface is depleted in the monoclinic phase. 
The values of both $\gamma$ and susceptibility continuously approach zero as the system approaches the structural phase boundary at $x_c$ = 0.035, suggesting that the transition approaches a second-order transition as the system approaches $x_c$. 
This enables us to consider the quantum criticality at the structural critical point $x_c$.

Using $\gamma$ = 6.3 mJ/molK$^2$ for $x$ = 0.04, we estimate the ratio of the specific heat jump at $T_c$ to $\gamma T_c$ to be $\Delta C_p(T_c)/\gamma T_c$ $\simeq$ 1.5, which is comparable to 1.43, predicted by the BCS weak coupling limit. 
From the magnetic field dependence of $\rho$, shown in Fig.~4(b), the upper critical field $H_{c2}(T)$ (defined as the midpoint of the resistivity transition) is obtained for Ir$_{1-x}$Pt$_x$Te$_2$ with $x$ = 0.04; the extrapolation to 0 K using the Werthamer-Helfand-Hohenberg (WHH) theory \cite{PhysRev.147.295} gives an $H_{c2}(0)$ value of 0.17 T, as shown in Fig.~4(a). 
We estimate the coherence length $\xi$ = $(\Phi_0/(2\pi H_{c2}(0)))^{1/2}$ to be 44 nm.
These results demonstrate that Ir$_{1-x}$Pt$_x$Te$_2$ is a type-II superconductor in the weak coupling limit. 

\begin{figure}
\begin{center}
\includegraphics[width=8cm]{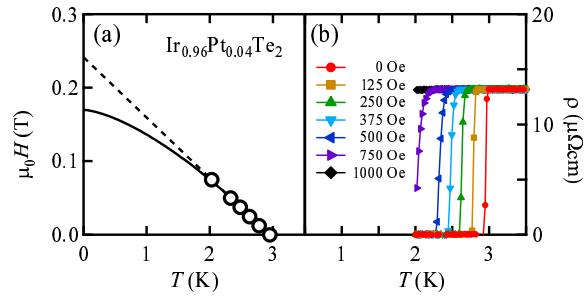}
\caption{\label{fig4}
(Color online)
(a) Temperature dependence of upper critical field $H_{c2}$ for Ir$_{1-x}$Pt$_x$Te$_2$ with $x$ = 0.04. The solid line shows the WHH behavior.
(b) Low-temperature resistivity in various applied fields for Ir$_{1-x}$Pt$_x$Te$_2$ with $x$ = 0.04. 
}
\end{center}
\end{figure}

The overall behavior of the system is summarized in the electronic phase diagram presented in Fig.~1(b). 
Using Pt doping as a system-controlling parameter, the structural phase transition in IrTe$_2$ is reduced in temperature, and a new superconducting phase emerges. 
Bulk superconductivity appears for $x$ $>$ 0.03, with a maximum $T_c$ of 3.1 K at $x$ = 0.03, then $T_c$ decreases with increasing $x$. 
Superconductivity disappears for $x$ $>$ 0.10.

Although the origin of the structural phase transition in IrTe$_2$ has not yet been determined, analogies between IrTe$_2$ and NaTiO$_2$ suggest that orbital ordering is likely to be involved in the transition of IrTe$_2$. 
NaTiO$_2$ and IrTe$_2$ consist of triangular lattices of transition-metal ions with a formal electron count of Ti$^{3+}$ (one electron in the $t_{2g}$ orbital) or Ir$^{4+}$ (one hole in the $t_{2g}$ orbital); thus, electron-hole symmetry exists between them. 
NaTiO$_2$ exhibits a phase transition at approximately 260 K, which is accompanied by structural distortion and a drop in the magnetic susceptibility \cite{JPSJ.61.2156, doi:10.1021/cm970538c}. 
The distortion is characterized by a uniform elongation of the triangular lattice along the $b$-axis in NaTiO$_2$, while a uniform shrinkage occurs in IrTe$_2$. 
Pen {\it et al.} theoretically proposed that orbital ordering occurs along the $b$-axis as well as the formation of one-dimensional $S$ = 1/2 chains \cite{PhysRevLett.78.1323}, which are consistent with the observed lattice distortion. 
This type of chain preferentially undergoes spin-singlet pair formation, consistent with the nonmagnetic ground state in NaTiO$_2$, although the structural modulation, indicative of a spin Peierls transition, has not yet been observed. 
Using this analogy and with the knowledge of electron-hole symmetry, we expect that orbital ordering similar to that in NaTiO$_2$ occurs in the phase transition of IrTe$_2$.

Quantum spin fluctuations often manifest themselves in the non-Fermi liquid behavior in itinerant magnets. 
In Ir$_{1-x}$Pt$_x$Te$_2$, the exponent $n$ of the temperature-dependent resistivity $\Delta\rho \propto T^n$ is estimated to be $n$ $\simeq$ 2.8 $\pm$ 0.1, and is almost independent of doping $x$. 
This exponential behavior has been observed, for example, in TiSe$_2$, and has been attributed to phonon-assisted interband scattering \cite{PhysRevLett.103.236401}. 
This process is very likely to occur in IrTe$_2$ because of the contribution of Ir $5d$ and Te $5p$ bands \cite{Soulard20052008}. 
We did not observe a noticeable change in the exponent $n$ at $x_c$, which is expected in the presence of quantum criticality \cite{PhysRevLett.103.236401}, although the peak in $\gamma$ and dip in the Debye temperature $\Theta_D$, indicative of phonon softening, shown in Figs.~1(c) and 1(d), respectively, suggest that critical fluctuations may exist around $x_c$. 
The appearance of superconductivity in a narrow range near $x_c$ is in accordance with the presence of such (presumably orbital) fluctuations, which are thought to mediate superconductivity.

In summary, we have found that the monoclinic phase of the triangular lattice of IrTe$_2$ is rapidly suppressed with Pt doping at the Ir site. 
Breaking of the Ir-Ir bonds, a characteristic of the monoclinic phase in pure IrTe$_2$, takes place across the critical concentration of $x_c$ $\simeq$ 0.035, and a superconducting phase arises. 
This finding opens up many new opportunities to realize superconductivity using nonmagnetic critical points.

\section*{Acknowledgment}
We would like to thank S. Watanabe and T. C. Kobayashi for valuable discussions. 
Part of this work was performed at the Advanced Science Research Center, Okayama University.
It was partially supported by a Grant-in-Aid for Young Scientists (B) (23740274) from Japan Society for the Promotion of Science (JSPS) and the Funding Program for World-Leading Innovative R\&D on Science and Technology (FIRST Program) from JSPS.


\

\textit{Note added in proof} - We noticed a paper by J. J. Yang \textit{et al}. [Phys. Rev. Lett. \textbf{108} (2012) 116402], reporting the superconductivity in Pd$_x$IrTe$_2$ and Ir$_{1-x}$Pd$_x$Te$_2$.

\end{document}